# A Space-Efficient Approach towards Distantly Homologous Protein Similarity Searches

Akash Nag*
Department of Computer Science
The University of Burdwan
Burdwan, West Bengal, India 713104
E-mail: nag.akash.cs@gmail.com

Sunil Karforma
Department of Computer Science
The University of Burdwan
Burdwan, West Bengal, India 713104
E-mail: sunilkarforma@yahoo.com

*Abstract:* Protein similarity searches are a routine job for molecular biologists where a query sequence of amino acids needs to be compared and ranked against an ever-growing database of proteins. All available algorithms in this field can be grouped into two categories – either solving the problem using sequence alignment through dynamic programming, or, employing certain heuristic measures to perform an initial screening followed by applying an optimal sequence alignment algorithm to the closest matching candidates. While the first approach suffers from huge time and space demands, the latter approach might miss some protein sequences which are distantly related to the query sequence. In this paper, we propose a heuristic pair-wise sequence alignment algorithm that can be efficiently employed for protein database searches for moderately sized databases. The proposed algorithm is sufficiently fast to be applicable to database searches for short query sequences, has constant auxiliary space requirements, produces good alignments, and is sensitive enough to return even distantly related protein chains that might be of interest.

## I. INTRODUCTION

During the past three decades, there has been a plethora of sequence alignment algorithms that have aimed at making the job of biologists easier by helping to identify homologies between two protein sequences, thus eventually establishing their phylogeny. Most of these algorithms can be divided into two categories based on the algorithmic approach adopted – those based on dynamic programming (DP); and those based on heuristics. Though sequence alignments are often quite useful, their primary usefulness stems from database searches. Often the purpose of a gene can be inferred from finding other genes (whose purpose is already known) which are similar to this gene, by searching through a genomic database [1]. The DP based approach is more accurate, but time consuming; and provides an optimal alignment between two sequences. However, they suffer from their huge demands of time and space, which often make them impractical for database searches. On the other hand, heuristic algorithms are blazingly fast and hence perfectly suited for database searches; however, heuristics do not produce optimal results and hence, often miss out on potential matches. In this paper, we propose a heuristic algorithm that strikes a balance between accuracy and rapid search, by reporting even distantly homologous sequences and finding out good alignments between two protein sequences. The algorithm is well suited for searching short query sequences in moderate sized databases like Swiss-Prot [2] within acceptable time limits. The novelty of the algorithm lies in its constant auxiliary space complexity.

## II. PRIOR WORK

The first dynamic programming algorithm for optimal global sequence alignment was presented by Needleman and Wunsch [3] in their landmark paper in 1970. They introduced the notions of gap penalties and alignment scores – critical concepts for developments that were to come later. Since then, there has been several developments in this regard, the most popular being an improvement of their work by Smith and Waterman [4] that led to local alignments rather than global alignments, which were biologically more relevant. Both these algorithms were optimal, i.e. they return the best possible alignment between two sequences, dependent on the gap penalties used. The primary drawbacks of the Smith-Waterman algorithm are its space and time complexity, the latter being $O(M^2N)$, with $M$ and $N$ being the lengths of the two protein sequences being aligned. Osamu Gotoh [5] improved the algorithm by reducing its time complexity down to $O(M.N)$, which, however, was still too slow to be effectively used for database searches. Altschul and Erickson [6] tweaked the algorithm further to incorporate affine gap costs which resulted in more meaningful alignments. They also suggested that the $M \times N$ dynamic programming matrix can be reduced to just two one-dimensional arrays to improve the space requirements of the algorithm. This latter idea was materialized by Myers and Miller [7] in 1988.

Despite the various improvements to the Smith-Waterman dynamic programming algorithm, its time complexity was too high for it to be successfully used for database searches. Despite this, the algorithm was actually used by biologists for years before finally the heuristic algorithms took over. The first popular heuristic algorithm to be successfully used for database search was the FASTA algorithm by Lipman and Pearson [8] [9]. Unlike the Smith-Waterman algorithm which was a sequence alignment algorithm, the FASTA algorithm was tailor-made for database searches. It reduced search times by leaps and bounds, and quickly became extremely popular with the biological community, and is still being used by biologists through the EMBL-EBI web service. However, the most important development came with the advent of BLAST [10] that revolutionized database search. BLAST was not only faster than all its predecessors, it was blazingly fast and for the first time, protein database search was no longer a waiting game. However, BLAST is effective in finding only ungapped alignments; hence gapped-BLAST [11] was designed, which is three times faster than its predecessor. A decade later, BLAT [12] came on the scene promising even faster alignments and searches, but BLAST continued to be the de facto database search tool within the community, primarily due to two limitations of BLAT: its requirement of a huge index; and its inability to accurately search sequences with lower than 80% match. An excellent comparison between the Smith-Waterman, FASTA and BLAST algorithms were presented by Pearson [13], and Shpaer et al. [14]; the latter conclusively showed that BLAST, though being the fastest is not accurate enough and often missed out on distant similarity between protein sequences.

## III. THE PROPOSED ALGORITHM

### A. Introduction

In this paper, we propose a randomized heuristic algorithm suitable for protein database searches that is extremely space-efficient and sensitive enough to detect even distant similarity between sequences.

The two primary advantages of our algorithm over BLAST are:

a) It has a constant auxiliary space complexity as compared to BLAST's $O(M.N)$ auxiliary space complexity.

b) It reports even distantly homologous proteins that BLAST misses out on.

Our algorithm is highly customizable, with intuitive parameters that can be tweaked for best performance. It is driven by the BLOSUM62 scoring matrix, but can be made to work with any other matrix as well by appropriately changing gap penalties in accordance.

### B. The Algorithm

The algorithm is primarily a pair-wise sequence alignment algorithm, and it will be discussed in the alignment context first. In Section B.3, we present the changes necessary for it to be adapted to database searches.

#### B.1 *Parameters*

Three types of gap penalties are used in this algorithm, which employs affine gap penalties:

a) **Periphery Gap Penalty (PGP):** This refers to all leading and trailing gaps, and can be set to 0 for local alignments or can be set appropriately for global and semi-global alignments.

b) **Gap Opening Penalty (GOP):** This refers to the introduction or start of a new gap run. It is generally set higher as indel events are rare in nature, and even when they do occur, short insertions or deletions in protein chains are rarer than larger ones.

c) **Gap Extension Penalty (GEP):** This refers to the extension of an already continuing gap run. Due to affine gap penalties being used, the algorithm penalizes short gap runs more than longer ones, in accordance with results seen in nature.

The algorithm takes four other parameters as defined below:

a) **Rounds:** The algorithm, being a randomized algorithm performs the same alignment several times, dictated by the Rounds parameter. E.g. If Rounds=10, each alignment is performed 10 times and the best scoring alignment is chosen from these 10 alignments. Increasing the number of rounds leads to better alignments at the cost of time.

b) **LFactor and SFactor:** The algorithm, in each step, takes a sub-sequence of the sequences and tries to align them. The length of the sub-sequence chosen is dependent on these two factors, where L and S refer to "large" and "small" respectively after the length of the two sequences (If the sequences are of equal size, the distinction is arbitrary). Both these factors range between 0 and 1. For optimum results, *SFactor* must be less than *LFactor* (e.g. *SFactor*=0.5, *LFactor*=1). A lower *SFactor* leads to selection of smaller portions of the smaller sequence, and a large *LFactor* leads to alignment with a maximal portion of the larger sequence.

c) **MinFactor:** It is the lower bound on the size of the sub-sequence chosen in each step, ranging between 0 and 1.

#### B.2 *Method*

The basic structure of the algorithm is presented in Fig. 1. The algorithm performs each alignment several times as dictated by the Rounds parameter. In each round, two random numbers, each between MinFactor and 1, are selected. These two values determine the sub-sequence size to be used for this round's alignment. The alignment is then performed using these parameters and the alignment with the highest score is returned.

The actual alignment has two sub-phases:
a. Chopping up the sequence into sub-sequences
b. Aligning each pair of sub-sequences (one from each sequence) against each other

The first phase is performed by the Align procedure in the algorithm, which randomly selects a subsequence from each sequence based on the parameters. These are then passed onto the subsequence alignment procedure. When the best alignment is returned, the actual subsequence used up in the alignment is then discarded. It is to be noted that it may so happen that the whole subsequence may not be actually used in an alignment stage. Only the portion of the sequence actually used, is deleted from the original sequence. The process continues till either of the sequences is exhausted, following which the other sequence is appended normally to get the aligned pair.

The subsequence alignment phase is performed by the getBestSubsequenceAlignment procedure in the algorithm. This is the heart of the algorithm where the actual alignment occurs. It slides one sequence against another sequence from one terminal to the other terminal with minimum overlap of 1 amino-acid. It then finds the alignment score (using the getVirtualAlignmentScore procedure) at each sliding-point, and then returns the shift at which the best score was obtained. The score is calculated based on the BLOSUM62 matrix. The process is illustrated in Fig. 2. Here we see two subsequences are being aligned with all possible shifts. The best shift occurs when the 2nd sub-sequence is shifted left 2 places.

After the best alignment is selected, the used and unused portions are determined. From Fig. 2 we see that the last two amino-acids do not contribute to the alignment and hence remain unused. Therefore, these are returned back to the original sequences, while the preceding amino-acids are deleted from the original sequence.

#### B.3 *Optimizations for use in database search*

The proposed algorithm in its presented form is too slow to be used in database search. Several optimizations are necessary for it to be successfully used for searching protein databases. The following are the optimizations applied when the algorithm is used for searching:

a) The Rounds parameter is set to 1, in the procedure AlignSequences

b) In the procedure *getBestAlignment*, the start and end parameters are set as follows: *start*=length(*small*)-1, *end*=length(*large*)-1. This has the effect of grading only those alignments in which the larger subsequence does not have any gaps.

#### B.4 *Peformance*

The novelty of the proposed algorithm is in its constant auxiliary space complexity, and hence is very suitable for implementations under critical memory constraints. Since this is a randomized algorithm, its time complexity is dependent on the parameters used. The choice of parameters shall dictate the subsequence size used for the actual shifting alignment phase. Consider a pair-wise alignment where the larger peptide consists of *M* amino-acids. Let the average sub-sequence size chosen be *L* and the number of rounds is *R*. Then, the total time taken by the algorithm is: $R.\left\{\frac{M}{L}.(2L^2)\right\}$. Hence, the time complexity is of the order of *O(R.M.L)*, and the time complexity for database search is *O(N.M.L)* where *N* is the total number of sequences in the database (and since R=1 when used for searching). A comparison between the complexities of the proposed algorithm with other well known sequence alignment algorithms, are presented in Table I.

B.5 *Scoring matrix, implementation and testing methodology*

The scoring matrix used in our algorithm was BLOSUM62. The protein database used for testing was UniProt-SwissProt containing 547,085 protein sequences in the standard FASTA file format. The algorithm was implemented in Java 8, under Windows/Intel-Dual-Core 2.3GHz.

## IV. RESULTS

The algorithm was tested for short query sequences up to 30 amino acids in length, with Periphery_Gap_Penalty=0, Gap_Opening_Penalty=10, Gap_Extension_Penalty=5, LFactor=0.5, SFactor=1, and MinFactor=0.5. The database contained 547,085 sequences. On average, a 30-amino-acid long peptide, when searched against the SwissProt database, with appropriately set threshold score, was completed in 197 seconds. Although the algorithm may seem drastically slow with respect to BLAST, it returned a large number of matches which BLAST missed when the threshold score was set appropriately. The algorithm is however several times faster than the Smith-Waterman algorithm.

Table I. Time Complexities of Sequence Alignment Algorithms

| Algorithm | Time Complexity | Auxiliary Space Complexity |
|---|---|---|
| Needleman-Wunsch | O(M.N) | O(M.N) |
| Smith-Waterman | O(M.N) | O(M.N) |
| Smith-Waterman (with Hirschberg improvement) | O(M.N) | O(N) |
| FASTA | O(M.N) | O(M.N) |
| BLAST | $O(M.N+N.20^W)$ | $O(M.N+20^W)$ |
| *Proposed algorithm* | *O(R.W.MAX(M,N))* | *O(1)* |

Presented above is a comparison between the time and space complexities of various sequence alignment algorithms for aligning two protein sequences consisting of M and N amino acids respectively [15]. W refers to the length of short sequences selected from the input sequence for alignment, and R refers to the value of the user-parameter ROUNDS.

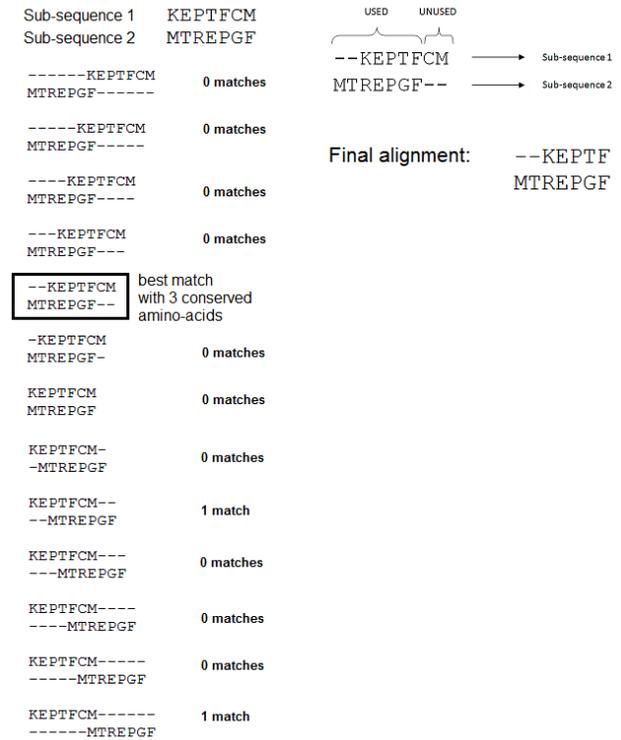

```
ALGORITHM AlignSequences(large, small)
begin
        for i=1 to Rounds do
        begin
                lf=MAX(minFactor, random());
                sf=MAX(minFactor, random());
                score=Align(large, small, lf, sf);
                if (score>bestScore) then bestScore=score;
        end
        return the alignment with score = bestScore;
end

PROCEDURE Align(large, small, LF, SF)
begin
        alignedLarge=alignedSmall=null;
        while (large is not empty AND small is not empty) do
        begin
                ls=length(large) * LF;
                ss=length(small) *SF * random();
                P=getBestSubsequenceAlignment(subsequence(large, ls), subsequence(small, ss));
                delete from large the used-up subsequence, and append it to alignedLarge;
                delete from small the first ss letters, and append it to alignedSmall;
        end
        append the remaining with dashes to align;
        return { alignedLarge, alignedSmall, SCORE(alignedLarge, alignedSmall) };
end

PROCEDURE getBestSubsequenceAlignment(large, small)
begin
        n=length(small)+length(large)-1;
        start=0, end=n-1;
        for i=start to end do
        begin
                k=i-length(small)+1;
                score=getVirtualAlignmentScore(large, small, k);
                bestScore=(score > bestScore ? score : bestScore);
        end
        return alignment with score = bestScore;
end
```

Figure 1. The Proposed Algorithm

Figure 2. A Sample Alignment

## V. CONCLUSIONS

The algorithm we presented in this paper is a big step forward for implementations requiring a low memory footprint. The algorithm in its present form is suitable for searching small databases only like SwissProt. With implementations in SIMD environments and further optimizations, we believe that this algorithm can be successfully used for searching large databases as well, especially when distant relationships need to be uncovered and where space constraints are acute. BLAST and gapped BLAST fail when searching distantly homologous sequences, and both are also heavy on memory requirements. And even though, the proposed algorithm is a heuristic one, the randomization adequately compensates for the lack in accuracy, as was evident during our sequence alignment tests with number of rounds set to 20. It also has a very small memory footprint, and is orders of magnitude faster than the Smith-Waterman algorithm. For largely homologous sequences however, BLAST is still the best algorithm out there for database searching.